\documentclass{article}
\usepackage{spconf,amsmath,graphicx}

\usepackage{enumitem}
\usepackage{amsfonts}
\setlist{nosep, leftmargin=14pt}

\usepackage{mwe} 

\newcommand\blfootnote[1]{%
  \begingroup
  \renewcommand\thefootnote{}\footnote{#1}%
  \addtocounter{footnote}{-1}%
  \endgroup
}

\title{Features fusion for Dual-View Mammography Mass Detection}

\name{Arina Varlamova$^{\star}$ \qquad Valery Belotsky$^{\star}$ \qquad Grigory Novikov$^{\star}$ \qquad Anton Konushin$^{\dagger}$ \qquad Evgeny Sidorov$^{\star \ddagger}$ }

\address{
    $^{\star}$ Third Opinion Platform;  $^{\dagger}$ Higher School of Economics;   $^{\ddagger}$ Lomonosov Moscow State University;
}
\begin{document}
%
\maketitle

\begin{abstract}
Detection of malignant lesions on mammography images is extremely important for early breast cancer diagnosis. In clinical practice, images are acquired from two different angles, and radiologists can fully utilize information from both views, simultaneously locating the same lesion. However, for automatic detection approaches such information fusion remains a challenge. In this paper, we propose a new model called MAMM-Net, which allows the processing of both mammography views simultaneously by sharing information not only on an object level, as seen in existing works, but also on a feature level. MAMM-Net's key component is the Fusion Layer, based on deformable attention and designed to increase detection precision while keeping high recall. Our experiments show superior performance on the public DDSM dataset compared to the previous state-of-the-art model, while introducing new helpful features such as lesion annotation on pixel-level and classification of lesions malignancy.
\end{abstract}
\blfootnote{~\copyright~2024 IEEE.  Personal use of this material is permitted.  Permission from IEEE must be obtained for all other uses, in any current or future media, including reprinting/republishing this material for advertising or promotional purposes, creating new collective works, for resale or redistribution to servers or lists, or reuse of any copyrighted component of this work in other works.}

\section{Introduction}
\label{sec:intro}
Breast cancer is one of the most common diseases, ranking first among the causes of cancer mortality among women\cite{sung2021global}. Approximately 12\% of all diagnosed cases of cancer in women are related to breast cancer, making it the dominant cancer-related disease in many countries.

Screening digital mammography is the most widely employed modality for the successful diagnosis of breast cancer\cite{dibden2020worldwide}. Presently, Artificial Intelligence (AI), mostly in the form of computer vision, is extensively utilized for tasks involving the automatic detection of breast cancer features, demonstrating high diagnostic accuracy comparable to or surpassing that of a radiologist\cite{yoon2023artificial}.

One of the primary tasks in forming radiological descriptions of mammography studies is the identification and comprehensive description of breast tissue abnormalities in both projections \cite{birads}. The differential diagnosis of lesions involves a comparative analysis of two views of each breast and between the two breasts. However, most current neural network architectures do not take into account the context of both views, limiting model performance.
To the best of our knowledge, existing two-view approaches fuse information  between different angles only on object level, using features obtained independently \cite{gnn, zhao2022check, ma2021cross}. In our study we propose new approach called \textbf{MAMM-Net}, where network is able to fuse information on features level in addition to object level, imitating radiologist's diagnosis process more naturally. We observe that additional information fusion helps model to effectively filter false positive detections while preserving the high recall, thus allowing to achieve state-of-the-art results.

\section{Related work}
\label{sec:related}
\subsection{Dual-view lesion segmentation}


The idea of dual-view segmentation was considered in several approaches. Liu {\textit{et al}. \cite{gnn} use a bipartite graph convolutional network to incorporate the intrinsic geometric and semantic relations of ipsilateral views. Ma {\textit{et al}. \cite{ma2021cross} propose a relation module to model correspondence between mass ROIs from different mammography images. In CL-Net \cite{zhao2022check} authors used cross-attention between object queries, generated by Deformable DETR \cite{deformabledetr} and a special module called Lesion Linker to verify object pairs.    

\subsection{Object recognition}
Transformer-based architectures gained popularity in tasks related to computer vision, including object detection and instance segmentation. Detection models originate from DETR \cite{detr}, which was the first to introduce the concept of query-based instance proposal. The idea was further adopted for instance segmentation \cite{m2f}, resulting in SOTA architecture at the time of writing.

\subsection{Multi-view segmentation}

3D segmentation in multi-camera space was actively explored in BEV Transformers \cite{li2022bevformer}. Feature fusion here was utilized through reprojection into the same 2D space and usage of deforamble attention in the local proximilty to points of interest. However, such approach is tricky in mammography application, since angles between images are in general unknown. In this settings there is a need for more flexible approach for generation of reference points, which can be achieved through deformable attention introduced in \cite{vit_defattn}.

\section{Methods}
\label{sec:methods}

\begin{figure}[t]
\centering
\centerline{\includegraphics[width=0.95\linewidth]{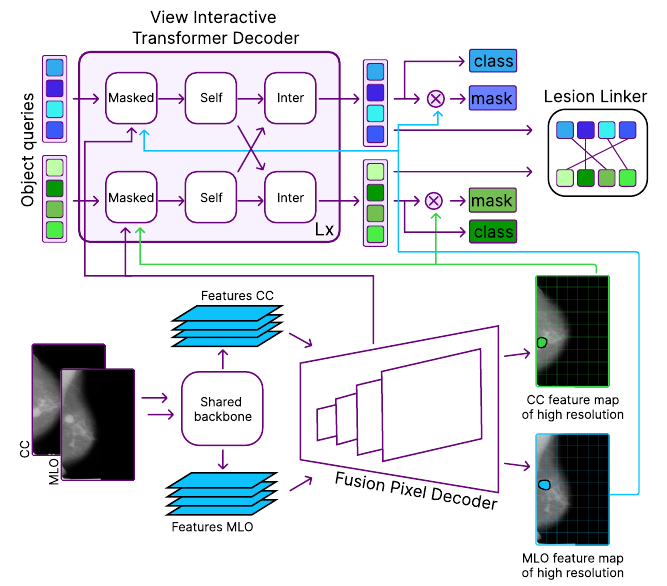}}
\caption{General overview of \textbf{MAMM-Net}: 1) Two different views are processed by a shared backbone independently; 2) Generated feature maps are processed by Fusion Pixel Decoder, which provides fused feature maps for View-Interactive Transformer Decoder's masked attention and feature maps of high resolution of both views for masks generation; 3) View-Interactive Transformer Decoder (VITD), consisting of blocks of masked-, self- and inter-attention, which outputs object queries, masks for both CC and MLO view, classification of found objects along with their malignancy scores; 4) Lesion Linker uses object queries from VITD to set correspondence between objects in CC and MLO views and outputs triplets of embeddings and pair classification.}
\label{fig:model_arch}
\end{figure}

Our proposed architecture will be explored further in this section, which we will reference as \textbf{MAMM-Net} (\textbf{M}ulti-view \textbf{A}ttention for \textbf{M}ass \textbf{M}atching). A brief overview of its structure is shown in Fig.~\ref{fig:model_arch}. 

We organize methods as follows: firstly, we briefly introduce key points of Mask2Former \cite{m2f} and CL-Net \cite{zhao2022check} architectures, which our model is based on, and then we explain our proposed Fusion Layer and View-Interactive Transformer Decoder (VITD) in more details.

\subsection{Mask2Former}

Our model mainly inherits the Mask2Former structure in its main components: backbone, pixel decoder, and transformer decoder with masked attention. The transformer decoder's key component is the masked attention operator, which allows the usage of spatial features restricted to the foreground area of the predicted masks. We left this part mostly unchanged except for using two object query branches for cranio-caudal (CC) and mediolateral oblique (MLO) mammography views and adding additional cross-attention logic between them.

Our Fusion Pixel Decoder, however, has more significant differences. Instead of generating feature maps independently, we combine them each time resolution increases. To provide intermediate feature maps of different resolutions, at each step we fuse features of CC and MLO projections into each other using our proposed Fusion Layers.  

\subsection{CL-Net}

Key components of CL-Net we were interested in are the VILD (View-Interactive Lesion Detector) and LL (Lesion Linker) modules. The main idea behind VILD is to add a cross-view inter-attention step in addition to self- and cross-attention in DETR's transformer decoder in order to help the model capture the relationship between objects in CC and MLO views. LL module at the same time aims to model correspondence between detected objects, outputing classification of the presence of object pair and link embeddings. 

\subsection{Fusion Pixel Decoder}

\begin{figure}[t]
\centering
\centerline{\includegraphics[width=0.85\linewidth]{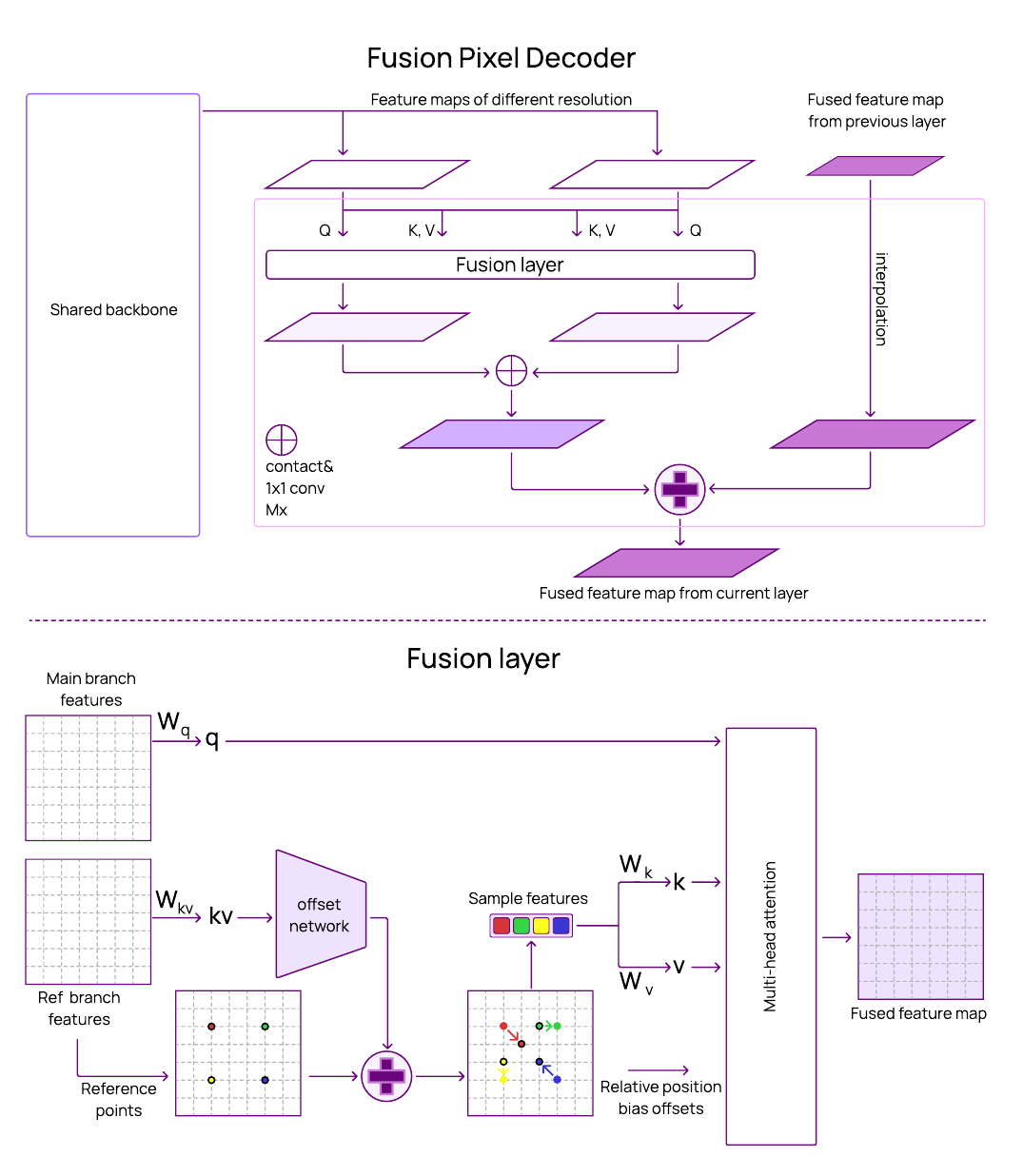}}
\caption{Architecture of Fusion Pixel Decoder (upper) and Fusion Layer (lower). 1) Fusion Pixel Decoder: The module uses feature maps of different resolutions for both CC and MLO views. Starting from the lowest resolution, feature maps are fused into each other using a special Fusion Layer and then are combined in a FPN manner. Fused feature maps of low resolution are transferred to the VITD to use in masked attention. The last fused feature map is used to generate masks of high resolution; 2) Fusion Layer: the main feature map (Q in Fusion Pixel Decoder) is used as queries in the multi-head attention module. Key and values are sampled from the reference feature map (K, V in Fusion Pixel Decoder). Generated queries, keys, and values are processed by a multi-head attention block.}\medskip
\label{fig:pixel_decoder}
\end{figure}

The Fusion Pixel Decoder is designed to provide both high-resolution mask feature maps for CC and MLO views as well as multi-scale fused feature maps for the VITD's masked attention. A brief overview of module structure is shown in Fig.~\ref{fig:pixel_decoder}. 

The module consists of several consequent blocks for combining information between projections. Each block takes two feature maps of different resolutions from the shared backbone, which are passed through the Fusion layer in two branches. Generated feature maps are further concatenated and convolved using ${1\times1}$ convolution to maintain the same channel dimension. Such feature maps are generated from low to high resolution and merged by element-wise addition in an FPN \cite{fpn} manner.

A key component of our pixel decoder is the Fusion layer, which is based on deformable attention \cite{vit_defattn}. The Fusion layer uses two feature maps: main to provide queries and reference from another view to provide key and values components. Since classic cross-attention between two feature maps is computationally demanding, it's essential to focus on a limited subset of spatial positions for key-values. In order to achieve that, we use the predefined amount of uniformly distributed points to initialize the start position of candidates. The special network generates their relative offsets, which are used to sample deformed key-values. Produced queries, keys, and values are then processed by a standard multi-head attention block.

It is worth noting that we use the Fusion Layer only in two blocks with feature maps with the lowest resolution. As resolution increases, we replace it with two independent $1 \times 1$ convolutions. Mask feature maps are generated by separately convolving the fused feature map with the highest resolution.

\subsection{VITD}

Following \cite{m2f}, blocks of our VITD consist of masked attention, self-attention, and FFN layers. We added an additional inter-attention layer to share object information between different views, similar to \cite{zhao2022check}. More formally, at $i^{th}$ decoder's iteration, module uses ${\boldsymbol{Q^{v}_{i}}=f_Q(\boldsymbol{X^{v}}_{i-1})}$ ${\in \mathbb{R}^{N \times C}}$ and ${\boldsymbol{K}_{i}}$, ${\boldsymbol{V}_{i}}$ ${\in \mathbb{R}^{H_iW_i\times C}}$ which are fused feature maps with applied transformations ${f_K(\cdot)}$ and ${f_V(\cdot)}$ respectively. ${\boldsymbol{X^{v}_{i}}}$ refers to query features at layer ${i}^{th}$ and specific view (CC or MLO). In these terms, output of the masked attention layer may be defined as follows:
\begin{equation}
    \boldsymbol{X_{m}}^{v}_i = softmax(\boldsymbol{\mathcal{M}^{v}}_{i-1} + \boldsymbol{Q^{v}}_i\boldsymbol{K}_{i}^{T})\boldsymbol{V}_{i} + \boldsymbol{X^{v}}_{i-1},
\end{equation}
where attention mask at spatial location $(x, y$) ${\boldsymbol{\mathcal{M}^{v}}_{i-1}(x, y) = 0}$ if the binarized output of the resized mask of corresponding view from a previous layer of decoder equals $1$, and $\infty$ otherwise. To utilize both high- and low-level features,  fused feature maps are fed to the decoder in a round-robin fashion.

After applying the self-attention layer to $\boldsymbol{X_{m}}^{v}_i$ for both CC and MLO, query vector is passed as $\boldsymbol{Q}$ to inter-attention layer with the same view, and as $\boldsymbol{K}$ and $\boldsymbol{V}$ to inter-attention layer for another projection. Outputs are processed by the feed-forward network, providing $\boldsymbol{X^v_i}$ for the next iteration of the decoder. 

Finally, we apply three feed-forward networks to each $\boldsymbol{X^v_i}$ to provide a classification of objects, their malignancy class, and masks embeddings ${\boldsymbol{E}^v_i}$, which are further multiplied to feature maps of high resolution from Fusion Pixel Decoder in order to produce objects masks. 

\subsection{Loss}
Our training loss is a combination of detection loss \cite{m2f}, linker loss \cite{zhao2022check} and malignancy loss. Detection and linker losses were implemented as in original publications, and a malignancy loss is a binary cross-entropy loss, where the target class is objects being malignant.

\section{Experiments}
\label{sec:experiments}

\subsection{Implementation Details}
We adopt EfficientNet-b3 \cite{tan2020efficientnet} as a backbone. We set the number of object queries and link queries to 100 and 50 respectively. In our experiments, we used 10 blocks in VITD. A downsample factor of 4 was used to generate reference points in Fusion layers. We set the number of heads of multi-head attention to 8 in both VITD and Fusion Pixel Decoder.

The network was trained with a batch size equal to 5, each item in a batch containing one mammography pair. We train our model with a learning rate set to $1 \times 10 ^{-4}$. To avoid overfitting we used L2-regularization with a weight of $1\times 10^{-5}$ and a variety of augmentations, including image flip, rotation, brightness and contrast modification, and random scaling.

\subsection{Datasets}

We conducted our experiments using the DDSM dataset \cite{ddscm1, ddscm2}. Originally, the dataset does not provide matching between objects in different views. In order to obtain it, cases were annotated by a skilled radiologist. In such annotations, we didn’t change any contour properties or coordinates, but rather made a classification of objects being pair or not.

\begin{figure}[t]
\centering
\centerline{\includegraphics[width=0.7\linewidth]{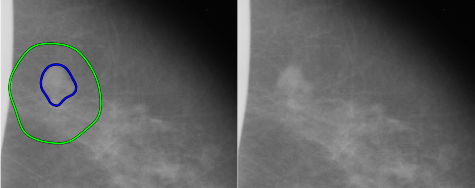}}
\caption{Example of prediction (blue) and ground true (green). Intersection over Union for those two objects equals 0.15 although contours clearly indicate the same object.
}\medskip
\label{fig:iou}
\end{figure}
Similar to \cite{ma2021cross, liu2020cross, zhao2022check, gnn}, we use recall ($R$) at $t$ false positives per image (FPI) to compare the performance of our model with other studies ($R@t$). Unlike in previous methods, we recall an object if it has IoU with ground true more than 0.1 since we have masks instead of bounding boxes and observe a relatively high proportion of true positives in diapason of $[0.1, 0.2]$. An example of such an object is shown in Fig.~\ref{fig:iou}.

Data splits are another matter of discussion. Train splits proposed in \cite{zhao2022check, liu2020cross, gnn} is ambiguous since Liu \textit{et al.} \cite{liu2020cross, gnn} mention 512 test cases and \cite{zhao2022check} doesn't specify test set size. At the same time, they both refer to \cite{campanini2004novel, ma2021cross}, which uses 512 images. We decided to follow the test split proposed in \cite{ma2021cross, campanini2004novel} with some modifications. Originally, images from cancer volumes without any masses were excluded. Since additional images without ground true objects can only worsen $R@t$, we selected all cases from these volumes for more representative comparison, resulting in 270 test cases and 1080 images.

\subsection{Comparison with other studies}

\begin{table}
 \caption{Comparison with previous SOTA on DDSM dataset \%. }
   \begin{center}
     {\small{
 \begin{tabular}{|c|c|c|c|}
 \hline
 Method & R@0.25 & R@0.5 & R@1.0 \\
 \hline
 Liu \textit{et al.} \cite{gnn} (AG-RCNN) &  &82.0&89.0\\
 \hline
 Zhao \textit{et al.} \cite{zhao2022check} (CL-Net) & 78.1 &83.1&88.0\\
 \hline
MAMM-Net (ours) & \textbf{81.6} & \textbf{87.9} & \textbf{90.6} \\
 \hline
 \end{tabular}
 }}
 \end{center}

 \label{table:comp1}
 \end{table}

 We show the comparison of our \textbf{MAMM-Net} with other methods in Table~\ref{table:comp1}. Result from Table~\ref{table:comp1} are reported from \cite{gnn, zhao2022check}. We keep the same FPIs as in \cite{zhao2022check} as the previous SOTA (CL-Net). It can be concluded that our model surpasses CL-Net by a large margin at all reported FPIs. It is worth noting, that we use a lighter backbone compared to ResNet-50 used in \cite{gnn, zhao2022check}, since we achived similar performance as CL-Net in setting with fusion on object level only. We show more details on that in section \ref{sec:vitd}. We believe that the result on $R@0.25$ is of the most significance and additionally provide binary malignancy metrics (per mammary gland) for this setting, such as ROC-AUC (85.3), sensitivity (80.2) and specificity (76.2). 

\subsection{Ablation Study}

\begin{table}
 \caption{Comparison different components of our model on DDSM dataset \%. }
   \begin{center}
     {\small{
 \begin{tabular}{|c|c|c|c|}
 \hline
 Method & R@0.25 & R@0.5 & R@1.0 \\
 \hline
  VITD&78.2  &83.3&87.3\\
 \hline
 Fusion Pixel Decoder&  77.9&80.1&85.9\\
 \hline
 MAMM-Net (ours) & \textbf{81.6} & \textbf{87.9} & \textbf{90.6} \\
 \hline
 \end{tabular}
 }}
 \end{center}

 \label{table:comp2}
 \end{table}
We evaluate the performance of the networks that use only one fusion component, either the VITD or the Fusion Pixel Decoder, against the performance of the \textbf{MAMM-Net}, which incorporates both fusion blocks. Table~\ref{table:comp2} illustrates the differences in recall for selected FPI thresholds. We further discuss the implementation details of networks with only one fusion component.

\subsubsection{VITD} 
\label{sec:vitd}

In this setting, we used a pixel decoder from Mask2Former , which outputs two independent sets of multi-scale feature maps for both views. We got similar values to CL-Net (78.2 vs 78.1, 83.3 vs 83.1, 87.3 vs 88.0 at respectively), which is to be expected since both models have similar key components. 

\subsubsection{Fusion Pixel Decoder}

In this setup the Fusion Pixel Decoder was left intact, while the VITD was significantly modified. Instead of two branches for different views, we used a single branch for object queries and excluded the inter-attention layer. For the mask in masked attention, we used a logical union of binarized mask outputs for both views. We skipped the LL block, using VITD's output directly in the loss computation and forcing matched objects to be predicted in the same position. Similar to \cite{zhao2022check}, we observe a significant drop in performance compared to the complete model. We support that forced prediction at the same position wasn't the best choice for modeling relationships between different objects. However, it was observed that for lower FPIs Fusion Pixel Decoder performs better than our full model (\textit{e.g.} 73.3 vs 69.7 $R@t0.14$). We hypothesize that Fusion Layers help the model to effectively filter candidates that look worthy of attention only from one view.

\section{Conclusion}
\label{sec:conclusion}

The main novelty introduced in our paper is the Fusion Layer, which enables feature-level fusion of two projections, leading to a decrease in false positive predictions without an increase in false negatives. This component is integrated into our newly proposed MAMM-Net architecture, designed for effective object recognition across two projections. Experiments on the DDSM dataset have demonstrated that our architecture outperforms previous state-of-the-art models.

\section{Acknowledgments}
\label{sec:acknowledgments}



No funding was received for conducting this study.

\section{Compliance with Ethical Standards}
\label{sec:compliance}

This research study was conducted retrospectively using human subject data sourced from DDSM. Ethical approval was not required as
confirmed by the license attached with the open access data.

\bibliographystyle{IEEEbib}
\bibliography{main}

\begin{thebibliography}{10}

\bibitem{sung2021global}
Hyuna Sung, Jacques Ferlay, Rebecca~L Siegel, Mathieu Laversanne, Isabelle Soerjomataram, Ahmedin Jemal, and Freddie Bray,
\newblock ``Global cancer statistics 2020: Globocan estimates of incidence and mortality worldwide for 36 cancers in 185 countries,''
\newblock {\em CA: a cancer journal for clinicians}, vol. 71, no. 3, pp. 209--249, 2021.

\bibitem{dibden2020worldwide}
Amanda Dibden, Judith Offman, Stephen~W Duffy, and Rhian Gabe,
\newblock ``Worldwide review and meta-analysis of cohort studies measuring the effect of mammography screening programmes on incidence-based breast cancer mortality,''
\newblock {\em Cancers}, vol. 12, no. 4, pp. 976, 2020.

\bibitem{yoon2023artificial}
Jung~Hyun Yoon, Kyungwha Han, Hee~Jung Suh, Ji~Hyun Youk, Si~Eun Lee, and Eun-Kyung Kim,
\newblock ``Artificial intelligence-based computer-assisted detection/diagnosis (ai-cad) for screening mammography: Outcomes of ai-cad in the mammographic interpretation workflow,''
\newblock {\em European Journal of Radiology Open}, vol. 11, pp. 100509, 2023.

\bibitem{birads}
EA~Sickles and Bassett~LW D’Orsi~CJ,
\newblock ``Acr bi-rads® mammography,''
\newblock {\em ACR BI-RADS® Atlas, Breast Imaging Reporting and Data System}, 2013.

\bibitem{gnn}
Yuhang Liu, Fandong Zhang, Chaoqi Chen, Siwen Wang, Yizhou Wang, and Yizhou Yu,
\newblock ``Act like a radiologist: towards reliable multi-view correspondence reasoning for mammogram mass detection,''
\newblock {\em IEEE Transactions on Pattern Analysis and Machine Intelligence}, vol. 44, no. 10, pp. 5947--5961, 2021.

\bibitem{zhao2022check}
Ziwei Zhao, Dong Wang, Yihong Chen, Ziteng Wang, and Liwei Wang,
\newblock ``Check and link: Pairwise lesion correspondence guides mammogram mass detection,''
\newblock in {\em European Conference on Computer Vision}. Springer, 2022, pp. 384--400.

\bibitem{ma2021cross}
Jiechao Ma, Xiang Li, Hongwei Li, Ruixuan Wang, Bjoern Menze, and Wei-Shi Zheng,
\newblock ``Cross-view relation networks for mammogram mass detection,''
\newblock in {\em 2020 25th International Conference on Pattern Recognition (ICPR)}. IEEE, 2021, pp. 8632--8638.

\bibitem{deformabledetr}
Xizhou Zhu, Weijie Su, Lewei Lu, Bin Li, Xiaogang Wang, and Jifeng Dai,
\newblock ``Deformable detr: Deformable transformers for end-to-end object detection,'' 2021.

\bibitem{detr}
Nicolas Carion, Francisco Massa, Gabriel Synnaeve, Nicolas Usunier, Alexander Kirillov, and Sergey Zagoruyko,
\newblock ``End-to-end object detection with transformers,'' 2020.

\bibitem{m2f}
Bowen Cheng, Ishan Misra, Alexander~G. Schwing, Alexander Kirillov, and Rohit Girdhar,
\newblock ``Masked-attention mask transformer for universal image segmentation,''
\newblock {\em CoRR}, vol. abs/2112.01527, 2021.

\bibitem{li2022bevformer}
Zhiqi Li, Wenhai Wang, Hongyang Li, Enze Xie, Chonghao Sima, Tong Lu, Qiao Yu, and Jifeng Dai,
\newblock ``Bevformer: Learning bird's-eye-view representation from multi-camera images via spatiotemporal transformers,'' 2022.

\bibitem{vit_defattn}
Zhuofan Xia, Xuran Pan, Shiji Song, Li~Erran Li, and Gao Huang,
\newblock ``Vision transformer with deformable attention,'' 2022.

\bibitem{fpn}
Tsung-Yi Lin, Piotr Dollár, Ross Girshick, Kaiming He, Bharath Hariharan, and Serge Belongie,
\newblock ``Feature pyramid networks for object detection,'' 2017.

\bibitem{tan2020efficientnet}
Mingxing Tan and Quoc~V. Le,
\newblock ``Efficientnet: Rethinking model scaling for convolutional neural networks,'' 2020.

\bibitem{ddscm1}
Michael Heath, Kevin Bowyer, Daniel Kopans, Richard Moore, and W.~Philip Kegelmeyer,
\newblock ``The digital database for screening mammography,''
\newblock {\em Proceedings of the Fifth International Workshop on Digital Mammography}, 2001.

\bibitem{ddscm2}
Michael Heath, Kevin Bowyer, Daniel Kopans, W.~Philip Kegelmeyer, Richard Moore, Kyong Chang, , and S.~MunishKumaran,
\newblock ``Current status of the digital database for screening mammography,''
\newblock {\em Digital Mammography}, 1998.

\bibitem{liu2020cross}
Yuhang Liu, Fandong Zhang, Qianyi Zhang, Siwen Wang, Yizhou Wang, and Yizhou Yu,
\newblock ``Cross-view correspondence reasoning based on bipartite graph convolutional network for mammogram mass detection,''
\newblock in {\em Proceedings of the IEEE/CVF Conference on Computer Vision and Pattern Recognition}, 2020, pp. 3812--3822.

\bibitem{campanini2004novel}
Renato Campanini, Danilo Dongiovanni, Emiro Iampieri, Nico Lanconelli, Matteo Masotti, Giuseppe Palermo, Alessandro Riccardi, and Matteo Roffilli,
\newblock ``A novel featureless approach to mass detection in digital mammograms based on support vector machines,''
\newblock {\em Physics in Medicine \& Biology}, vol. 49, no. 6, pp. 961, 2004.

\end{thebibliography}
\end{document}